\documentclass[twocolumn,showpacs,aps,prl,superscriptaddress,preprintnumbers,amsmath,amssymb,floatfix]{revtex4}

\usepackage{graphicx}
\usepackage{dcolumn}
\usepackage{bm}

\usepackage{relsize}
\usepackage{xspace}
\def\babar{\mbox{\slshape B\kern-0.1em{\smaller A}\kern-0.1em
    B\kern-0.1em{\smaller A\kern-0.2em R}}}
\def\pep2{PEP-II}

\def\Km{K^-}
\def\Kp{K^+}
\def\piz{\pi^0}
\def\pip{\pi^+}

\def\DsTT{D_{sJ}^*(2317)^+}
\def\DsTO{D_s^{*}(2112)^+}
\def\Kbar{\kern 0.2em\overline{\kern -0.2em K}{}\xspace}

\newcommand{\gevc}{\ensuremath{{\mathrm{\,Ge\kern -0.1em V\!/}c}}\xspace}
\newcommand{\mevc}{\ensuremath{{\mathrm{\,Me\kern -0.1em V\!/}c}}\xspace}
\newcommand{\gevcc}{\ensuremath{{\mathrm{\,Ge\kern -0.1em V\!/}c^2}}\xspace}
\newcommand{\mevcc}{\ensuremath{{\mathrm{\,Me\kern -0.1em V\!/}c^2}}\xspace}

\newcommand{\BABARPubYear}    {03}
\newcommand{\BABARPubNumber}  {011}
\newcommand{\SLACPubNumber} {9711}

\begin{document}

\preprint{BABAR-PUB-\BABARPubYear/\BABARPubNumber}
\preprint{SLAC-PUB-\SLACPubNumber}

\title{\boldmath Observation of a Narrow Meson Decaying to
$D_s^+\piz$ at a Mass of 2.32~\gevcc}

%
\author{B.~Aubert}
\author{R.~Barate}
\author{D.~Boutigny}
\author{J.-M.~Gaillard}
\author{A.~Hicheur}
\author{Y.~Karyotakis}
\author{J.~P.~Lees}
\author{P.~Robbe}
\author{V.~Tisserand}
\author{A.~Zghiche}
\affiliation{Laboratoire de Physique des Particules, F-74941 Annecy-le-Vieux, France }
% ...
\collaboration{The \babar\ Collaboration}
\noaffiliation

\date{\today}

\begin{abstract}
We have observed a narrow state near 2.32~\gevcc
in the inclusive $D_s^+ \piz$ invariant mass distribution
from $e^+e^-$ annihilation data at energies near 10.6~GeV. The observed width
is consistent with the experimental resolution. The small
intrinsic width and the quantum numbers of the final state indicate
that the decay violates isospin conservation.
The state has natural spin-parity and the low mass
suggests a $J^P=0^+$ assignment. 
The data sample corresponds to an
integrated luminosity of 91~${\rm fb}^{-1}$
recorded by the \babar\  detector at the \pep2 asymmetric-energy $e^+e^-$
storage ring.
\end{abstract}

\maketitle

We have found a narrow state decaying to
$D_s^+\piz$ at a mass near 2.32~\gevcc.  This result is obtained
from a 91~${\rm fb}^{-1}$ data sample recorded both on and off the
$\Upsilon(4S)$ resonance by the \babar\ detector at the \pep2
asymmetric-energy $e^+e^-$ storage ring.  

Experimental information on the spectrum of the $c\overline{s}$ meson states
is limited. The $^1\!S_0$ ground state, the
$D_s^+$ meson, is well-established, as is the $^3\!S_1$ ground state,
the $\DsTO$. Only two other $c\overline{s}$ states have been observed thus
far~\cite{Hagiwara:2002pw}. The $D_{s1}(2536)^+$ has been detected in
its $D^* K$ decay mode and analysis of the $D^*$ decay angular
distribution prefers $J^P = 1^+$~\cite{Alexander:1993nq}.  
The $D_{sJ}^*(2573)^+$ was
discovered in its $D^0 \Kp$ decay mode and so has natural spin-parity.
The assignment $J^P=2^+$ is consistent with the data, but is not
established~\cite{Kubota:1994gn}.  

The spectroscopy of $c\overline{s}$ states is simple in the limit of
large charm-quark mass~\cite{DeRujula:1976kk,Isgur:1991wq}.  
In that limit,
the total angular momentum $\vec{j}=\vec{l}+\vec{s}$ of the light
quark, obtained by summing its orbital and spin angular momenta,
is conserved.  The $P$-wave states, all of which have
positive parity, then have $j=3/2$ or $j=1/2$.  Combined with the
spin of the heavy quark, the former gives total angular momentum $J=2$
and $J=1$, while the latter gives $J=1$ and $J=0$.  
The $J^P=2^+$ and $J^P=1^+$ members of the $j=3/2$ doublet are
expected to have small width~\cite{Godfrey:1991wj}, 
and are identified with the
$D_{sJ}^*(2573)^+$ and $D_{s1}(2536)^+$, respectively, although the latter 
may include a small admixture of the $j=1/2$, $J^P = 1^+$ state.
Theoretical models typically predict masses between
2.4 and 2.6~\gevcc for the remaining two
states~\cite{Godfrey:1985xj,Godfrey:1991wj,DiPierro:2001uu}, both of
which should decay by kaon emission.
They would be expected to have large 
widths~\cite{Godfrey:1991wj,DiPierro:2001uu}
and hence should be difficult to detect. 

The experimental and theoretical status of the
$P$-wave $c\overline{s}$ states thus can be summarized by stating
that experiment has provided good candidates for the two states that
theory predicts should be readily observable, but has no candidates
for the two states that should be difficult to observe because
of their large predicted widths.

The \babar\ detector is
a general purpose, solenoidal, magnetic spectrometer, which is
described in detail elsewhere \cite{Aubert:2001tu}. The detector 
components employed in this analysis are discussed briefly here.
Charged particles are detected
and their momenta measured by a combination of a cylindrical drift chamber (DCH)
and a silicon vertex tracker (SVT), both operating within a
1.5-T solenoidal magnetic field. 
A ring-imaging Cherenkov detector (DIRC) is used for
charged-particle identification. Electrons are identified
and photons measured with a CsI electromagnetic calorimeter (EMC).

The objective of this analysis is to investigate the inclusively-produced
$D_s^+ \piz$ mass spectrum by combining charged particles
corresponding to the decay $D_s^+\to \Kp \Km \pip$~\footnote{The 
inclusion of charge-conjugate 
configurations is implied throughout this letter.} with $\piz$ candidates
reconstructed from a pair of photons. 
Events of interest are required to
contain at least three
reconstructed tracks yielding a net charge of $\pm 1$ and at least two
photons each of which must have energy greater than 100~MeV,
and to have a ratio of the second to the zeroth 
Fox-Wolfram moment~\cite{Fox:1978vu} less than 0.9.
Charged-kaon candidates are
selected based on the Cherenkov-photon information from the DIRC
together with the measured energy loss in the SVT and DCH.

A $\Kp \Km$ candidate pair is combined with a third track that
fails the kaon criteria (and so is treated as a pion) in a geometrical
fit to a common vertex. An acceptable $\Kp\Km\pip$ candidate must have a fit
probability greater than 0.1\% and a trajectory consistent with 
originating from the $e^+e^-$ 
luminous region.
Background from $D^0 \to \Kp\Km$, which is evident from the
corresponding $\Kp\Km$ mass distribution, is removed by requiring
that the $\Kp\Km$ mass be less than 1.84~\gevcc. 

A candidate $\piz$ is formed by constraining a photon pair to emanate from
the intersection of the $\Kp \Km \pip$ candidate trajectory
and the beam envelope, 
performing a one-constraint fit to the
$\piz$ mass, and requiring a fit probability 
greater than 1\%.  A given event
may yield several acceptable $\piz$ candidates. We retain only
those candidates for which neither 
photon belongs to another acceptable $\piz$ candidate.

Finally, to reduce combinatorial background from
the continuum and eliminate background from $B$-meson decay, each
$\Kp \Km \pip \piz$ candidate must have a momentum $p^*$ in 
the $e^+e^-$ center-of-mass frame greater than 2.5~\gevc.

\begin{figure}
\vskip -0.15in
\includegraphics[width=\linewidth]{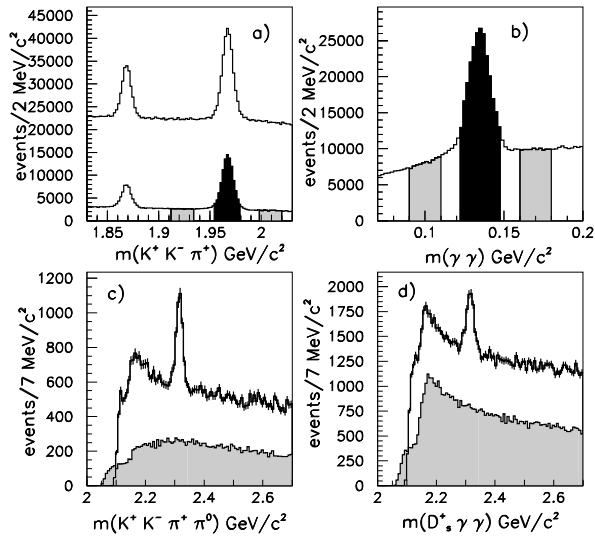}
\caption{\label{fig:kkpi}
(a) The distribution of $\Kp\Km\pip$ mass for all candidate events.
Additional selection criteria, described in the text, have been
used to produce the lower histogram.
(b) The two-photon mass distribution from $D_s^+\piz$ candidate events.
$D_s^+$ and $\piz$ signal and sideband regions are shaded.
(c) The $D_s^+\piz$ mass distribution for
candidates in the $D_s^+$ signal (top histogram) and $\Kp\Km\pip$
sideband regions (shaded histogram) of (a).
(d) The $D_s^+\gamma\gamma$ mass distribution for signal $D_s^+$
candidates and a photon pair from the $\piz$ signal region of (b) 
(top histogram) and
the sideband regions of (b) (shaded histogram). 
}
\end{figure}

The upper histogram in Fig.~\ref{fig:kkpi}(a) shows the $\Kp\Km\pip$ mass 
distribution for all candidates. Clear peaks corresponding to $D^+$ and $D_s^+$
mesons are seen. To reduce the background further,
only those candidates with $\Kp\Km$ mass
within 10~\mevcc of the $\phi(1020)$ mass
or with $\Km\pip$ mass within 50~\mevcc of the $\Kbar^*(892)$
mass are retained; these densely populated
regions in the $D_s^+$ Dalitz plot do not overlap.
The decay products of the vector particles $\phi(1020)$ and 
$\Kbar^*(892)$ exhibit the
expected $\cos^2 \theta_h$ behavior required
by conservation of angular momentum, where $\theta_h$ is the
helicity angle. The signal-to-background ratio
is further improved by requiring $|\cos\theta_h|>0.5$. The lower histogram
of Fig.~\ref{fig:kkpi}(a) shows the net effect of these additional selection
criteria. The $D_s^+$ signal ($1.955<m(\Kp\Km\pip)<1.979$~\gevcc)
and sideband ($1.912<m(\Kp\Km\pip)<1.934~\gevcc$ and
$1.998<m(\Kp\Km\pip)<2.020$~\gevcc) regions are shaded.
The $D_s^+$ signal peak, consisting of approximately 80,000 events,
is centered at a mass of $(1967.20 \pm 0.03)$~\mevcc
(statistical error only).

Figure~\ref{fig:kkpi}(b) shows the mass distribution for all
two-photon combinations associated
with the selected events.  The $\piz$ signal ($122<m(\gamma\gamma)<148$~\mevcc)
and sideband ($90<m(\gamma\gamma)<110$~\mevcc 
and $160<m(\gamma\gamma)<180$~\mevcc)
regions are shaded.
Candidates in the $D_s^+$ signal region of Fig.~\ref{fig:kkpi}(a) are
combined with the 
mass-constrained $\piz$ candidates to yield the mass distribution
of Fig.~\ref{fig:kkpi}(c). A clear,
narrow signal at a mass near 2.32~\gevcc is seen.  The shaded histogram
represents the events in the $D_s^+\to\Kp\Km\pip$ mass sidebands combined with 
the $\piz$ candidates. 
In Fig.~\ref{fig:kkpi}(d)
the mass distributions result from the combination of the $D_s^+$
candidates with the photon pairs from the $\piz$ signal and 
sideband regions of Fig.~\ref{fig:kkpi}(b) (the sideband
distribution is again shaded). In this case, all photon pairs in
the signal region of Fig.~\ref{fig:kkpi}(b) are used.
In Figs.~\ref{fig:kkpi}(c) and \ref{fig:kkpi}(d)
the 2.32~\gevcc signal is absent from the sideband distributions
indicating quite
clearly that the peak is associated with the $D_s^+\piz$ system.  
No other signal
in the region up to 2.7~\gevcc is evident in these plots, except
for a small $\DsTO \to D_s^+\piz$ signal in Fig.~\ref{fig:kkpi}(c).

In order to improve mass resolution, the nominal
$D_s^+$ mass~\cite{Hagiwara:2002pw} has been used to calculate the
$D_s^+$ energy for the distributions of Fig.~\ref{fig:kkpi}(d), for the
$D_s^+$ signal distribution of Fig.~\ref{fig:kkpi}(c),
and for all subsequent mass distributions involving
$D_s^+$ candidates.

\begin{figure}
\vskip -0.15in
\includegraphics[width=0.75\linewidth]{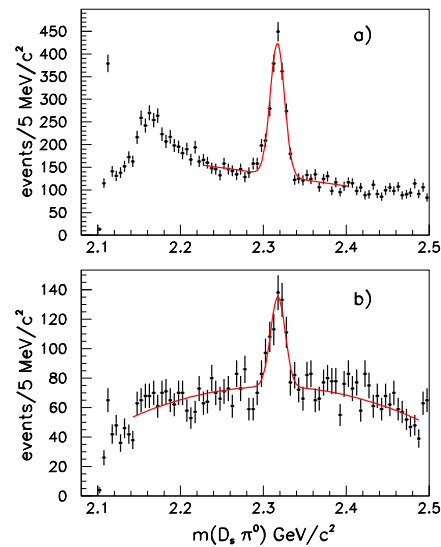}
\vskip -0.15in
\caption{\label{fig:dspiz} 
The $D_s^+\piz$ mass distribution
for (a) the decay $D_s^+  \to \Kp \Km \pip$
and (b) the decay $D_s^+  \to \Kp \Km \pip \piz$.
The fits to the mass distributions as described in the
text are indicated by the curves.
}
\end{figure}

The $D_s^+\piz$ mass distribution for $p^*(D_s^+\piz)>3.5$~\gevc
is shown in Fig.~\ref{fig:dspiz}(a). 
Similar distributions produced for $p^*$ values ranging from 2.5 to
4.5~\gevc show the same prominent peak at the same mass value.
The fit function drawn on Fig.~\ref{fig:dspiz}(a) comprises a
Gaussian function describing the $2.32$~\gevcc signal and 
a third-order polynomial background distribution function. 
The fit yields $1267 \pm 53$ candidates in the signal Gaussian with 
mass $(2316.8\pm 0.4)$~\mevcc and standard deviation
$(8.6\pm 0.4)$~\mevcc (statistical errors only).
The systematic uncertainty in the mass is conservatively estimated
to be less than 3~\mevcc.
The broad peak in Fig.~\ref{fig:dspiz}(a)
centered at 2.16~\gevcc is due to random $\DsTO\gamma$ combinations where
$\DsTO \to D_s^+\gamma$.

The signal, which we label $\DsTT$, is observed in both
the $\phi\pip$ and $\Kbar^{*0}\Kp$ decay modes of the $D_s^+$.
In addition, a sample of $D_s^+ \to \Kp\Km\pip\piz$ decays is
selected by adding $\piz$ candidates (refit to the $\Kp\Km\pip$ vertex)
to each $\Kp\Km\pip$ candidate.
The purity of this $D_s^+$ sample is enhanced
by requiring a $\piz$ fit probability of at least 10\% and
selecting the $K^{*\pm}$, $\Kbar^{*0}$,
$\phi$, or $\rho^+$ mass regions for the relevant two-body subsystems. 
Each resulting $D_s^+$ candidate is combined with a second $\piz$ candidate
with lab momentum greater than 300~\mevc. 
A clear $\DsTT$ signal is observed as shown in 
Fig.~\ref{fig:dspiz}(b). A Gaussian fit
yields $273 \pm 33$ events with a mean of $(2317.6\pm 1.3)$~\mevcc
and width $(8.8\pm 1.1)$~\mevcc (statistical errors only).
The mean and width are consistent with
the values obtained for the $D_s^+ \to \Kp\Km\pip$ decay mode. 
The mass distribution of the $D_s^+\to \Kp\Km\pip\piz$ sample (not shown) 
peaks at $(1967.4 \pm 0.2)$~\mevcc (statistical error only).

We use a Monte Carlo simulation to investigate the possibility that 
the $\DsTT$ signal could 
be due to reflection from other charmed states.
This simulation includes $e^+e^-\to c\bar c$ events 
and all known charm states and decays. The generated events were
processed by a detailed 
detector simulation and subjected to the same reconstruction
and event-selection procedure as that used for the data. No
peak is found in the 2.32~\gevcc $D_s^+\piz$ signal region.
In addition, no signal peak is produced when
the $K^{\pm}$ and $\pi^{\pm}$ identities are deliberately exchanged.

Mass resolution estimates for the $\Kp\Km\pip\piz$ system are
obtained directly from the data using a fit to the mass distribution
$D_s^+\to \Kp\Km\pip\piz$. The measured width from this
mode is consistent
with that of the $\DsTT$ signal. A simulation of the
$\DsTT$ decay to $\Kp\Km\pip\piz$ yields a similar
mass resolution after event reconstruction and selection criteria
have been satisfied. We conclude that the
intrinsic width of the $\DsTT$ is small ($\Gamma \lesssim 10$~MeV).

The $\cos \theta_h$ distribution of the $\DsTT$ decay with
respect to its direction in the $e^+e^-$ center-of-mass frame 
has been investigated.
The efficiency-corrected distribution is consistent with being flat,
as expected for a spin-zero particle, or
for a particle of higher spin that is produced unpolarized.

\begin{figure}
\includegraphics[width=0.8\linewidth]{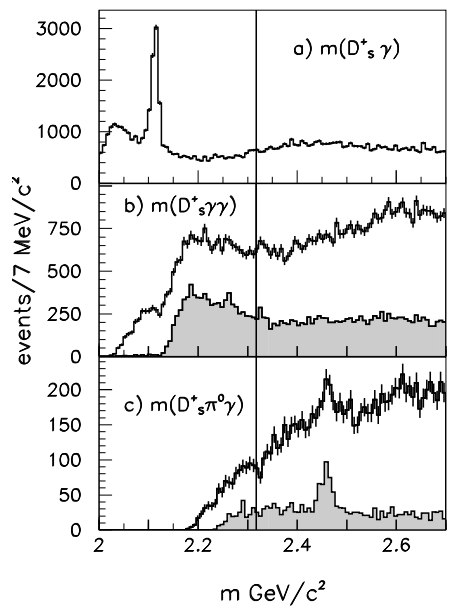}
\vskip -0.1in
\caption{\label{fig:dsgammas}
The mass distribution for (a)~$D_s^+\gamma$ and (b)~$D_s^+\gamma\gamma$
after excluding photons from the signal region of Fig.~\ref{fig:kkpi}(b).
(c)~The $D_s^+\piz\gamma$ mass distribution.
The lower histograms of (b) and (c) correspond to $D_s^+\gamma$
masses that fall in the $\DsTO$ signal region as described in
the text. The vertical line indicates the $\DsTT$ mass.
}
\end{figure}

We have also performed a search for the decay 
$\DsTT \to D_s^+\gamma$.  
Shown in Fig.~\ref{fig:dsgammas}(a) is the $D_s^+\gamma$ mass 
distribution obtained by combining a $D_s^+$ candidate
in the signal region of Fig.~\ref{fig:kkpi}(a) with a photon
with an energy of at least 150~MeV
that does not belong to a $\gamma\gamma$ combination in the
signal region of Fig.~\ref{fig:kkpi}(b). The requirement
that the $p^*$ of the $D_s^+\gamma$ system be greater than
3.5~\gevc is also imposed. There is a clear $\DsTO$ signal,
but no indication of $\DsTT$ production.

The $D_s^+ \gamma\gamma$ mass distribution for 
$p^*(D_s^+ \gamma\gamma) > 3.5$~\gevc,
excluding any photon that belongs to the $\piz$ signal
region of Fig.~\ref{fig:kkpi}(b),
is shown as the upper histogram of 
Fig.~\ref{fig:dsgammas}(b). 
No signal is observed near 2.32~\gevcc. The shaded histogram
corresponds to the subset of combinations for which either $D_s^+ \gamma$
combination lies in the $\DsTO$ region, defined as 
$2.096 < m(D_s^+ \gamma) < 2.128$~\gevcc. Again, no $\DsTT$ signal 
is evident, thus demonstrating
the absence of a $\DsTO \gamma$ decay mode at the present level of
statistics. 

The $D_s^+ \piz \gamma$ mass distribution,
excluding any photon that belongs to any $\piz$ candidate,
is shown as the upper histogram of 
Fig.~\ref{fig:dsgammas}(c). 
The shaded histogram corresponds to the subset of combinations
in which the $D_s^+ \gamma$ mass falls in the $\DsTO$ region.
No signal is observed near 2.32~\gevcc in either case.
A small peak, however, is visible near a mass of 2.46~\gevcc. 
This mass corresponds to the
overlap region of the $\DsTO \to D_s^+ \gamma$ and 
$\DsTT \to D_s^+ \piz$ signal bands
that, because of the small widths of both the $\DsTO$
and $\DsTT$ mesons, produces a narrow peak in the
$D_s^+ \piz \gamma$ mass distribution that survives
a $\DsTO$ selection.

If the peak in the $D_s^+ \piz \gamma$ mass 
distribution of Fig.~\ref{fig:dsgammas}(c) were
due to the production of a narrow state with mass near 2.46~\gevcc
decaying to $\DsTO \piz$, the kinematics are such that a peak would
be produced in the  $D_s^+\piz$ mass distribution at a mass
near 2.32~\gevcc. Such a $D_s^+\piz$ mass peak, however, would
have a root-mean-square of $\sim 15$~\mevcc, which is significantly larger
than that obtained for the $\DsTT$ signal.
In addition, Monte Carlo studies indicate that if the apparent
signal at 2.46~\gevcc were due to a state that decays entirely
to $\DsTO\piz$, it would produce only one-sixth the signal
we observe at 2.32~\gevcc.

Although we rule out the decay of a state of mass 2.46~\gevcc
as the sole source of the $D_s^+\piz$ mass peak corresponding to
the $\DsTT$, such a state may be produced in addition
to the $\DsTT$. However, the complexity of the overlapping kinematics 
of the $\DsTO \to D_s^+\gamma$ and $\DsTT \to D_s^+ \piz$ 
decays requires more detailed study,
currently underway, in order to arrive at a definitive conclusion.

The decay of any $c\overline{s}$ state to $D_s^+\piz$
violates isospin conservation,
thus guaranteeing a small width.
It is possible that the decay proceeds via $\eta-\piz$ mixing, 
as discussed by Cho and Wise~\cite{Cho:1994zu}.
For a parity-conserving decay only a spin-parity
assignment in the natural $J^P$
series $\{0^+, 1^-, 2^+, \dots\}$ is allowed.
The low mass compared to those of the 
$D_{s1}(2536)^+$ and the $D_{sJ}^{*}(2573)^+$ favors $J^P = 0^+$. 
In this case, decay to $D_s^+ \gamma$ is
excluded. However, decay of the $\DsTT$ to $D_s^{*}(2112)^+\gamma$
is allowed and might compete with decay by pion emission. 
The shaded mass distribution of
Fig.~\ref{fig:dsgammas}(b) suggests that this mode is absent, 
at least at the present level of statistics. 
This may simply indicate that decay by pion emission is 
favored over radiative decay.

Further studies are under way. If, however, the tentative $J^P=0^+$
assignment is confirmed, the low mass, small width, and decay mode of the 
$\DsTT$ are quite different from those
predicted by potential 
models~\cite{Godfrey:1985xj,Godfrey:1991wj,DiPierro:2001uu}. 

In summary, in 91~${\rm fb}^{-1}$ of data collected
by the \babar\  experiment 
we have observed a narrow state in the inclusive $D_s^+\piz$
mass distribution near 2.32~\gevcc. 
We find no evidence for the decay of this state to
$D_s^+\gamma$, $\DsTO\gamma$, or $D_s^+\gamma\gamma$.
Since a $c\bar{s}$ meson of this mass contradicts
current models of charm meson 
spectroscopy~\cite{Godfrey:1985xj,Godfrey:1991wj,DiPierro:2001uu}, 
either these models need modification
or the observed state is of a different type altogether,
such as a four-quark state.

\begin{acknowledgments}
We are grateful for the excellent luminosity and machine conditions
provided by our \pep2\ colleagues, 
and for the substantial dedicated effort from
the computing organizations that support \babar.
The collaborating institutions wish to thank 
SLAC for its support and kind hospitality. 
This work is supported by
DOE
and NSF (USA),
NSERC (Canada),
IHEP (China),
CEA and
CNRS-IN2P3
(France),
BMBF and DFG
(Germany),
INFN (Italy),
FOM (The Netherlands),
NFR (Norway),
MIST (Russia), and
PPARC (United Kingdom). 
Individuals have received support from the 
A.~P.~Sloan Foundation, 
Research Corporation,
and Alexander von Humboldt Foundation.

\end{acknowledgments}

\bibliography{note616}

\end{document}